# Hypersingular Integral Equations and Applications to Porous Elastic Materials

**Gerardo Iovane**[1], **Michele Ciarletta**[2]

[1,2]*Dipartimento di Ingegneria dell'Informazione e Matematica Applicata, Universita' di Salerno, Italy*

In this paper a treatment of hypersingular integral equations, which have relevant applications in many problems of wave dynamics, elasticity and fluid mechanics with mixed boundary conditions, is presented. The main goal of the present work is the development of an efficient direct numerical collocation method. The second part of the work is devoted to the application of the method to the porous elastic materials. This part deals with classical problem for cracks dislocated in a certain very specific porous elastic material, described by a Cowin-Nunziato model. By applying Fourier integral transform the problem is reduced to some integral equations. For the plane-strain problem we operate with a direct numerical treatment of a hypersingular integral equation. We also study stress-concentration factor, and investigate its behaviour versus porosity of the material. At the end we investigate the possibility to extend the results in the thermoelastic case.

## 1 introduction

There are a number of theories about mechanical properties of porous materials. Cowin and Nunziato proposed a theory to describe properties of homogeneous elastic materials with voids free of fluid [1]. This theory is a special case of microstretch elasticity of Eringen, when micropolar effects are discarded. Moreover, the theory of Cowin and Nunziato is more appropriated then other theories for the study of special continuum and geological materials, like rocks, soils, and manufactured porous materials like ceramics and pressed powders. More in detail, it is used for solid material with voids, but without any other phase, like liquid or gas.

Generally, this theory is founded on the balance of energy, where presence of the pores involves additional degree of freedom, namely, the fraction of elementary volume. As a consequence, the bulk mass density is given by the product of two fields, the void volume fraction and the mass density of the matrix (elastic) material. An exact explicit solution is well known in the problem for a crack dislocated in the classical linear elastic space [2]. If a normal load is applied to the faces of the crack, then the shape of the faces near the edge of the crack under this stress can be represented explicitly as a root-square function. This permits analytical calculation of the stress concentration coefficient in the classical case.

Obviously, stress concentration analysis is also very important in engineering practice for porous materials. Therefore, the main goal of the present work is to construct a strict solution of the static crack problem for the plane-strain cracks dislocated in the linear elastic granular (porous) space.

Starting from the basic equations, we demonstrate the applicability of the Fourier transform that allows us to reduce the problem to some integral equations and to construct a direct numerical collocation technique to solve this equation. In the plane-strain problem we treat numerically a certain hypersingular integral equation, and a special kind of the collocation technique can be also applied to such equation. Finally, we show in the figure how the calculated stress concentration coefficient at the crack edge depends upon variation of some physical and geometric parameters.

## 2 Mathematical tools for Elastic Media with Voids

Let us consider elastic material with voids which possesses a reference configuration with a constant volume fraction $\nu_0$. The considered theory asserts that the constant mass density $\rho$ has the decomposition [1] $\rho=\gamma\nu$, where $\gamma$ is the density of the matrix material, and $\nu$ ($0<\nu\leq1$) is the volume fraction field. Let $\phi$ ($\phi=\nu-\nu_0$) be the change in volume fraction from the reference one. Then the linear theory of homogeneous and isotropic elastic material with voids is described by the following system of partial differential equations

$$\begin{cases} \mu\Delta\bar{u} + (\lambda+\mu)\,grad\,div\bar{u} + \beta grad\phi = 0 \\ \alpha\Delta\phi - \xi\phi - \beta div\bar{u} = 0 \end{cases} \quad (1)$$

where $\mu$ and $\lambda$ are classical elastic constants; $\alpha$, $\beta$ and $\xi$ - some constants related to porosity of the medium. Besides, $\bar{u}$ denotes the displacement vector. Obviously, if $\beta=0$, then the elastic and the ``porosity'' fields are independent. Thus, in the case $\beta=0$ the stress-strain state is insensitive to the function $\phi$. The components of the stress tensor are defined, in terms of the functions $\bar{u}$ and $\phi$, by the following relations ($\delta_{ij}$ is the Kronecker's delta)

$$\begin{cases} \sigma_{ij} = \lambda\delta_{ij}\varepsilon_{kk} + 2\mu\varepsilon_{ij} + \beta\phi\delta_{ij} \\ \varepsilon_{ij} = \frac{1}{2}(u_{i,j} + u_{j,i}) \end{cases} \quad (2)$$

---

[1] Research Scientist, iovane@diima.unisa.it
[2] Professor, ciarlett@diima.unisa.it

Let us formulate the plane-strain boundary value problem for the porous (granular) medium. In this case $\bar{u} = \{u_x(x,y), u_y(x,y), 0\}$, and the basic system (1) can be rewritten as follows

$$\frac{\partial^2 u_x}{\partial x^2} + c^2\frac{\partial^2 u_x}{\partial y^2} + (1-c^2)\frac{\partial^2 u_y}{\partial x \partial y} + H\frac{\partial\phi}{\partial x} = 0$$

$$\frac{\partial^2 u_y}{\partial y^2} + c^2\frac{\partial^2 u_y}{\partial x^2} + (1-c^2)\frac{\partial^2 u_x}{\partial x \partial y} + H\frac{\partial\phi}{\partial y} = 0 \quad (3)$$

$$l_1^2\left(\frac{\partial^2\phi}{\partial x^2} + \frac{\partial^2\phi}{\partial y^2}\right) - \frac{l_1^2}{l_2^2}\phi - \left(\frac{\partial u_x}{\partial x} + \frac{\partial u_y}{\partial y}\right) = 0$$

with

$$\frac{\sigma_{xx}}{\lambda+2\mu} = \frac{\partial u_x}{\partial x} + (1-2c^2)\frac{\partial u_y}{\partial y} + H\phi,$$

$$\frac{\sigma_{yy}}{\lambda+2\mu} = (1-2c^2)\frac{\partial u_x}{\partial x} + \frac{\partial u_y}{\partial y} + H\phi, \quad (4)$$

$$\frac{\sigma_{xy}}{\mu} = \frac{\partial u_x}{\partial x} + \frac{\partial u_y}{\partial y}.$$

where

$$c^2 = \frac{\mu}{\lambda+2\mu}, \quad H = \frac{\beta}{\lambda+2\mu},$$

$$l_1^2 = \frac{\alpha}{\beta}, \quad l_2^2 = \frac{\alpha}{\xi}, \quad (5)$$

where the first two numbers $c$, $H$ are dimensionless and the quantities $l_1$, $l_2$ have the dimension of length.

Let us consider a thin crack of the length $2a$ with plane faces, dislocated over the segment $-a<x<a$ along the x-axis. Let the plane-strain deformation of this crack be caused by a constant stress $\sigma_{yy}=\sigma_0$ applied at infinity (i.e. at $y=\pm\infty$). Then, due to linearity of the problem, it is easily seen that both the shape of the crack's faces and the stress concentration at the crack's edges are the same as in the problem with a solution decaying at infinity and the following boundary conditions corresponding to the case when the normal load $-\sigma_0$ is symmetrically applied to the faces of the crack and there is no load at infinity. For the last problem the boundary conditions over the line $y=0$ are

$$\sigma_{xy} = 0, \quad \frac{\partial\phi}{\partial y} = 0 \quad (|x|<\infty),$$

$$\sigma_{yy} = -\sigma_0 \quad (|x|<a), \quad u_y = 0 \quad (|x|>a). \quad (6)$$

By applying a rather standard approach based on Fourier Transform along x axis the boundary value problem can be reduced to the hypersingular integral equation

$$\int_{-a}^{a} g(\xi)K(x-\xi)d\xi = -(1-N)^2\frac{\sigma_0}{2\mu}, \quad |x|<a \quad (7)$$

where $g$ is the opening of the crack face, $N = (l_2^2/l_1^2)H$ ($0 \leq N < 1$) is the coupling number and

$$K(x) = \frac{1}{2\pi}\int_{-\infty}^{\infty}\frac{s}{q(s)}\Big[2Nc^2s^2(q-|s|) + (1-N)$$

$$(1-N-c^2)q\Big]e^{-isx}ds = \frac{1}{\pi}\int_{0}^{\infty}L(s)\cos(sx)ds, \quad (8)$$

with

$$q = q(s) = \sqrt{s^2+1-N},$$

$$L(s)\frac{s}{q(s)}\Big[2Nc^2s^2(q-s) + (1-N)(1-N-c^2)q\Big] \quad (9)$$

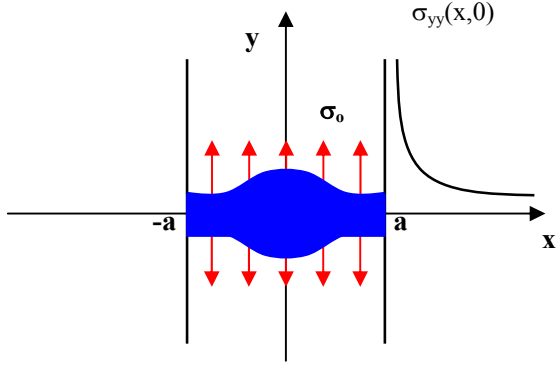

**Figure 1:** A scheme of the stressed crack.

### 3 Analytical and Numerical implementation for the Plane Problem

Let us start from the evident estimate

$$q(s) \approx s\left[1 + \frac{1-N}{2s^2} + O\left(\frac{1}{s^4}\right)\right], \quad s \to \infty \quad (10)$$

So the asymptotic behavior of the symbolic function is

$$L(s) \approx (1-N)^2(1-c^2)s + O\left(\frac{1}{s}\right), \quad s \to \infty \quad (11)$$

and so by substituting into the integral (8) we obtain

$$K(x) \approx -\frac{(1-N)^2(1-c^2)}{\pi x^2} + O(\ln|x|), \quad x \to 0 \quad (12)$$

So the kernel is hypersingular indeed.
The numerical method that we apply to solve the hypersingular integral equation has been proposed in our previous paper [3]. It is based on extraction of a characteristic hypersingular part of the kernel. Let us represent the full equation as follows

$$\int_{-a}^{a}\left[\frac{1}{(x-t)^2} + K_*(x,t)\right]g(t)dt = f'(x), \quad (13)$$

with $x \in (-a,a)$, where the regular part $K_*$ may admit a weak singularity.
The stress concentration factor k can be obtained starting from the relation

$$\frac{(1-N)^2}{2\mu}\sigma_{yy}(x,0) = \int_{-a}^{a} g(\xi)K(x-\xi)d\xi \quad (14)$$

with

$$\sigma_{yy}(x,0) = \begin{cases} \sigma_0, & |x| < a \\ \dfrac{k}{\sqrt{x^2 - a^2}}, & |x| > a \end{cases} \quad (15)$$

Then

$$k = \lim_{x \to a+0} \frac{|\sigma_{yy}(x,0)|}{\mu a(1-c^2)}\sqrt{x^2 - a^2} = $$
$$= \frac{2}{\pi}\lim_{x \to a+0}\left|\int_{-a}^{a}\frac{g(\xi)d\xi}{(x-\xi)^2}\right|\sqrt{x^2 - a^2} \quad (16)$$

where the last limit is numerically evaluated and displayed in the following figure.

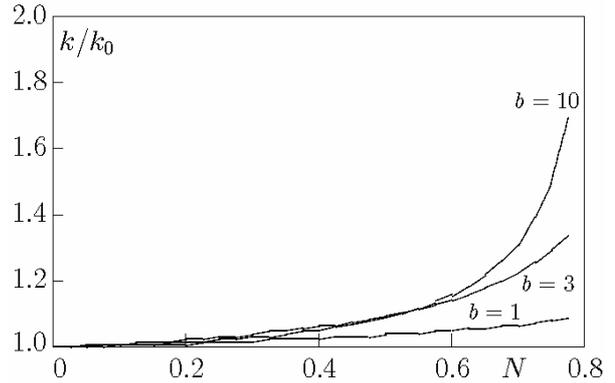

**Figure 2:** Relative value of the concentration factor $k$ with respect to $k_o$ (in the classical elastic medium) versus coupling number, plane linear cracks: $c^2=0.2$

### 4 From the elastic to the thermoelastic point of view

In this section we consider the field equations from a thermoelastic point of view. In this case the equivalent equations of (1) are

$$\begin{cases} \mu\Delta\bar{u} + (\lambda+\mu)grad\,div\bar{u} + \beta grad\phi + \\ -bgrad\theta = 0 \\ \alpha\Delta\phi - \xi\phi - \beta div\bar{u} + m\theta = 0 \\ \Delta\theta = 0 \end{cases} \quad (17)$$

In the 2-dimensional plane problem the previous equations can be written as

$$\frac{\partial^2 u_x}{\partial x^2} + c^2 \frac{\partial^2 u_x}{\partial y^2} + (1-c^2)\frac{\partial^2 u_y}{\partial x \partial y} + H\frac{\partial \phi}{\partial x} +$$

$$- B\frac{\partial \theta}{\partial x} = 0$$

$$\frac{\partial^2 u_y}{\partial y^2} + c^2 \frac{\partial^2 u_y}{\partial x^2} + (1-c^2)\frac{\partial^2 u_x}{\partial x \partial y} + H\frac{\partial \phi}{\partial y} +$$

$$- B\frac{\partial \theta}{\partial y} = 0 \qquad (18)$$

$$l_1^2\left(\frac{\partial^2 \phi}{\partial x^2} + \frac{\partial^2 \phi}{\partial y^2}\right) - \frac{l_1^2}{l_2^2}\phi - \left(\frac{\partial u_x}{\partial x} + \frac{\partial u_y}{\partial y}\right) +$$

$$+ \frac{l_1^2}{l_3^2}\theta = 0$$

$$\Delta \theta = 0,$$

where

$$c^2 = \frac{\mu}{\lambda + 2\mu}, \quad H = \frac{\beta}{\lambda + 2\mu}, \quad B = \frac{b}{\lambda + 2\mu}$$

$$l_1^2 = \frac{\alpha}{\beta}, \quad l_2^2 = \frac{\alpha}{\xi}, \quad l_3^2 = \frac{\alpha}{m} \qquad (19)$$

By following the same approach of the elastic case thanks to the Fourier transform the problem can be solved. In particular the idea is to solve the last equation and then to substitute the solution in the other equation. The most important question is how to formulate correctly the boundary conditions for $\theta$. The most natural way is to consider the flow of the temperature equal to zero for $y=0, |x|>a$. Inside the crack ($y=0, |x|<a$) there are different possibilities:

1) Given flow of temperature

$$\left.\frac{\partial \theta}{\partial x}\right|_{y=0} = \left.\frac{\partial \theta}{\partial y}\right|_{x=0} = \begin{cases} f_0(x), & |x|<a \\ 0, & |x|>a \end{cases} \qquad (20)$$

2) Given temperature

$$\left.\theta\right|_{y=0} f_0(x), \quad |x|<a \qquad (21)$$

In this case we have a mixed-value boundary problem also for the temperature. As first approach it is difficult to combine this with other boundary conditions. Then let us continue with (20). By following the same strategy of the first part of the paper it easy to obtain

$$\sigma_0 = \left.\frac{\sigma_{yy}}{\lambda + 2\mu}\right|_{y=0} = \qquad (22)$$

$$= (1-c^2)\frac{\partial u_x}{\partial x} + \frac{\partial u_y}{\partial y} + H\phi - B\theta$$

This gives the main integral equation with respect to the function $g(t)$, $|x|<a$ like as in the elastic case. The results of this problem are quite similar to the elastic case and the plot of figure 2.

## 5 Conclusion and Perspectives

In this paper we have proposed a method based upon a reducing of stress concentration problem for cracks to some integral equations. In the plane-strain problem this is a hypersingular integral equation, which permits efficient direct numerical treatment.

In some cases the equations admit exact analytical solution in explicit form. The first case is for N=0, that is a classical linear elastic material. In this case the plane-strain problem's integral equation has only the characteristic component and the regular part K* vanishes. In the other cases we found numerical solution thanks to our numerical methods tested and fine-tuned on the previous case.

By investigating the influence of the porosity to the stress concentration factor shown in figure, we can discover, this factor in the medium with voids is always higher, under the same conditions, than in the classical elastic medium made of material of the skeleton. Further, as can be seen, influence of the porosity becomes more significant for larger cracks. From a thermoelastic point of view the framework appear complex respect to the combination of the boundary condition. For a given flow of the temperature the problem appear similar to the elastic case.

## References

[1] S.C.Cowin, J.W.Nunziato, Linear elastic materials with voids, Journ. Elasticity 13, 125-147, 1983.

[2] I.N.Sneddon, Mixed Boundary Value Problems in Potential Theory, North-Holland, Amsterdam, 1966.

[3] G.Iovane, I.K.Lifanov, M.A. Sumbatyan, On direct numerical treatment of hypersingular integral equations arising in mechanics and acoustics, Acta Mechanica, 2003 (accepted).